# Charge-Transfer Hyperbolic Polaritons in α-MoO$_3$/graphene heterostructures

J. Shen[1†], M. Chen[1†], V. Korostelev[1], H. Kim[2], P. Fathi-Hafshejani[3], M. Mahjouri-Samani[3], K. Klyukin[1]*, G-H. Lee[2], S. Dai[1]*

[1]*Materials Research and Education Center, Department of Mechanical Engineering, Auburn University, Auburn, Alabama 36849, USA*

[2]*Department of Materials Science and Engineering, Seoul National University, Gwanak-ro 1, Gwanak-gu, Seoul, 08826 Republic of Korea*

[3]*Department of Electrical and Computer Engineering, Auburn University, Auburn, Alabama 36849, USA*

[†]These authors contribute equally

*Correspondence to: sdai@auburn.edu and klyukin@auburn.edu

**Abstract**

Charge transfer is a fundamental interface process that can be harnessed for light detection, photovoltaics, and photosynthesis. Recently, charge transfer was exploited in nanophotonics to alter plasmon polaritons by involving additional non-polaritonic materials to activate the charge transfer. Yet, direct charge transfer between polaritonic materials hasn't been demonstrated. We report the direct charge transfer in pure polaritonic van der Waals (vdW) heterostructures of α-$MoO_3$/graphene. We extracted the Fermi energy of 0.6 eV for graphene by infrared nano-imaging of charge transfer hyperbolic polaritons in the vdW heterostructure. This unusually high Fermi energy is attributed to the charge transfer between graphene and α-$MoO_3$. Moreover, we have observed charge transfer hyperbolic polaritons in multiple energy-momentum dispersion branches with a wavelength elongation of up to 150%. With support from the DFT calculation, we find that the charge transfer between graphene and α-$MoO_3$, absent in mechanically assembled vdW heterostructures, is attributed to the relatively pristine heterointerface preserved in the epitaxially grown vdW heterostructure. The direct charge transfer and charge transfer hyperbolic polaritons demonstrated in our work hold great promise for developing nano-optical circuits, computational devices, communication systems, and light and energy manipulation devices.

Charge transfer (CT)—electrons or holes relocate to different regions or nearby materials—is a fundamental interface process that leads to important applications. CT facilitates efficient electron-hole separation for light detection[1] and photovoltaics[2, 3]. Electron or hole relocation also affects chemical reactions and can be exploited for photosynthesis[4, 5] and catalysis[6]. Recently, CT was investigated in van der Waals (vdW) materials where high-quality heterointerfaces can form regardless of lattice mismatching and host tightly bound electron-hole pairs called interlayer excitons[7-11]. In addition, CT from adjacent vdW layers metalizes graphene without electrostatic gating or chemical doping, thus allowing the alteration of surface plasmon polaritons—referred to as charge transfer plasmon polaritons (CPPs)[12, 13]. Current CCPs in graphene rely on involved non-polaritonic materials[12, 13], such as α-RuCl$_3$ and WO$_x$, to activate the CT inside the heterostructures. Yet, direct CT in polaritonic heterostructures consisting solely of polaritonic materials has not been achieved.

In this work, we report charge transfer hyperbolic polaritons (CHPs) based on direct CT between two polaritonic vdW materials—α-MoO$_3$ and graphene—in epitaxial α-MoO$_3$/graphene heterostructures. Despite a large difference in the work functions, the direct CT between α-MoO$_3$ and graphene is not evident in mechanically assembled α-MoO$_3$/graphene heterostructures[13]. The direct CT between α-MoO$_3$ and graphene observed in our work is attributed to the relatively pristine α-MoO$_3$/graphene interface[14-16] preserved by the vdW epitaxy. In our experiments, the CT elongates the wavelength λ of hyperbolic polaritons (HPs) from bare α-MoO$_3$ up to Δλ ~ 150%. Notably, we observed CHPs at various energy-momentum ($\omega$-$k$) dispersion branches of the α-MoO$_3$/graphene heterostructure with varying wavelength elongation (Δλ$_p$) at different branches. The CHPs measured by infrared (IR) nano-imaging are supported by Density Functional Theory (DFT) calculations. These two approaches consistently show that the Fermi energy ($E_F$) of graphene is ~ 0.6 eV due to the direct CT between polaritonic α-MoO$_3$ and graphene. Our work shows that the CT polaritonic interfaces are beneficial for various nanophotonic and thermal functionalities.

To investigate direct CT and CHPs at α-MoO$_3$/graphene heterointerface, we used IR nano-imaging with scattering-type scanning near-field optical microscopy (s-SNOM, Figure 1a). The α-MoO$_3$/graphene heterostructures were prepared by epitaxial growth[14, 15, 17] of α-MoO$_3$ crystals on exfoliated monolayer graphene. A representative device in Figure 1b shows thin α-MoO$_3$ on monolayer graphene with two thick single-crystal α-MoO$_3$ domains. The high-quality crystalline of the heterostructures was verified by polarized Raman spectroscopy (Supplementary S1). The characterization tool s-SNOM is an illuminated atomic force microscopy (AFM) that can simultaneously provide topography and nano-optical images of the underneath samples (Figure 1a). In the experiment, a sharp AFM tip acts as an antenna[18] to record the s-SNOM near-field amplitude $S(\omega)$ with a spatial resolution of ~ 10 nm. On polaritonic materials, the tip bridges the momentum mismatch and transfers energy between free-space light (wavelength λ$_0$ and frequency $\omega = 1/\lambda_0$) and confined light-matter waves—polaritons[19-24]. Therefore, various types of polaritons can be launched and detected by the AFM tip in s-SNOM experiments.

A representative s-SNOM image of the α-MoO$_3$/graphene heterostructure is plotted in Figure 1b. The imaging frequency $\omega$ = 870 cm$^{-1}$ falls inside the Restrahlen band of α-MoO$_3$, where its permittivity tensor components are ε$_{[100]}$ < 0, ε$_{[001]}$ > 0, and ε$_{[010]}$ > 0. Therefore, in-plane HPs in α-MoO$_3$ can be imaged by the s-SNOM[13, 25-29] as parallel fringes close to crystal edges along the [001] direction on the two α-MoO$_3$ slabs in Figure 1b. These fringes are standing wave interferences between the tip-launched and edge-reflected HPs. Therefore, the strongest oscillation is detected in proximity to the edges of α-MoO$_3$, while weakly damped oscillations extend into the interior of the α-MoO$_3$ crystal. In addition, the fringes reveal the superposition of HPs with multiple wavelengths λ$_p$ due to the hyperbolic response in α-MoO$_3$ (ε$_{[100]}$ε$_{[010]}$ < 0). Away from the α-MoO$_3$ slabs, the graphene covered by thin α-MoO$_3$ also exhibits weak fringes near the edge of the graphene, which mainly originates from the plasmonic response in graphene, as detailed later in the manuscript.

The *ω-k* dispersion of HPs imaged in our experiments clearly shows deviation from HPs in bare α-MoO$_3$ slabs. Figure 2a shows the s-SNOM line profiles extracted from the cyan-dotted line on the α-MoO$_3$ slab of Figure 1b. The polariton fringes were revealed as *S*(ω) oscillations as a function of the distance to the crystal edge (*L*). As the frequency (*ω*) increases, the polariton fringes move toward the edge (*L* = 0). The polariton momentum ($k = 2\pi/\lambda_p$) can be obtained by the Fourier Transform (FT) analysis of the s-SNOM line profiles. The FT spectra of the s-SNOM line profiles (Figure 2b) show a series of resonances marked with ■ and ●. These resonances correspond to the HPs at various *ω-k* dispersion branches of hyperbolic α-MoO$_3$ (Figure 2c). The resonance peak positions in the FT spectra of Figure 2b are plotted in Figure 2c (pink squares). Note that at ω = 860–880 cm$^{-1}$, weak ● resonances appear, with *k* around half of the second (from the left) ■ resonances. Therefore, these two types of resonances are treated as edge-launched and tip-launched HPs[30] from the same dispersion branch, respectively. Figure 2c shows that the *ω-k* dispersion of the α-MoO$_3$/graphene heterostructure deviates from the HP dispersion of the bare α-MoO$_3$ slab on SiO$_2$ without graphene (black dashed curves). However, it fits well with the calculation results (false color, see Supplementary S2 for details) by involving the underneath graphene with the Fermi Energy $E_F$ = 0.6 eV.

The unusual polariton dispersion and abnormally high Fermi Energy in α-MoO$_3$/graphene heterostructure are attributed to the direct CT between α-MoO$_3$ and graphene that results in the formation of the CHPs. Specifically, the CT metalizes graphene with strong plasmonic responses. The surface plasmon polaritons in metalized graphene hybridize with hyperbolic phonon polaritons in α-MoO$_3$, leading to their wavelength-elongated[31-33] hybrid polaritons—CHPs. To confirm the origin of the CHPs, we performed DFT calculations of the projected density of states (DOS, Figure 3a) in the α-MoO$_3$/graphene heterostructure. As previously reported, the α-MoO$_3$ contains a number of oxygen vacancies[17]. So, we implement oxygen vacancies (the absence of O$_1$, Figure 3c inset) caused by lattice mismatch between graphene and α-MoO$_3$ during the epitaxial growth. The DFT calculations show a strong CT in the epitaxial α-MoO$_3$/graphene heterostructure (Figure 3b). The electrons are transferred from the graphene to α-MoO$_3$ and accumulated on the Mo and O atoms, forming a two-dimensional (2D) electron gas confined in the bottom α-MoO$_3$ layer and leaving graphene p-doped (Figure 3a and b). Furthermore, the oxygen vacancies lead to gap states (Figure 3a) and a decrease in the dipole moment on the α-MoO$_3$ (010) surface. Therefore, the work function (6.9 eV)[34, 35] of α-MoO$_3$ decreases closer to that of graphene (4.6 eV)[12, 36]. The calculated shift of the Fermi energy is ~ 0.6 eV (Figure 3a), and the CT is $3.5 \times 10^{13}$ e$^-$/cm$^2$, in good agreement with our s-SNOM data. Note that $E_F$ = 0.6 eV is substantially higher than the $E_F$ of pristine graphene or graphene with unintentional environmental doping[37]—revealing the strong CT in our α-MoO$_3$/graphene heterostructures.

Although we observed the direct CT in epitaxial α-MoO$_3$/graphene heterostructures, this phenomenon is not evident in mechanically assembled α-MoO$_3$/graphene heterostructures[13]. The negligible CT in the latter can be attributed to the abundance of oxygen vacancies[38-41] and adsorbed moisture[41, 42] on the α-MoO$_3$ surface due to the air exposure during the fabrication. The abundant vacancies and adsorbates produce substantial gap states that significantly lower the work function of α-MoO$_3$ down to 5.35 eV[14, 38, 43, 44] and largely weaken the CT. Indeed, our DFT results (Figure 3c) show the detrimental effects of vacancies and adsorbates on the CT: the number of transferred electrons decreases with the increasing vacancies and adsorbates. Ideally, the defect- and adsorbates-free α-MoO$_3$/graphene heterointerfaces are predicted to support the strongest CT by our DFT calculations (Figure 3c). In vdW epitaxy, while the lattice mismatch between α-MoO$_3$ and graphene can also introduce oxygen vacancies and gap states, the α-MoO$_3$/graphene interface is adsorbates-free[16], leading to significantly weaker CT reduction. Therefore, evident CT can still be observed in our epitaxial α-MoO$_3$/graphene heterostructure.

After confirming the CT origin of the polaritonic responses in the α-MoO$_3$/graphene heterostructure, we provide additional analysis of the CHPs. Notably, the CT evidently elongates the HP wavelength from bare α-MoO$_3$. The wavelength elongation $\Delta\lambda = (\lambda_{CHP} - \lambda_{HP}) / \lambda_{HP}$ ($\lambda_{HP}$ is the polariton wavelength from bare α-MoO$_3$) reaches 150% in our experiments and even 350% in our calculations at higher ω (Figure 4). In addition, the wavelength elongation $\Delta\lambda$ varies at different dispersion branches ($l$ = 0, 1, 2, …): $\Delta\lambda$ decreases at the increasing branch index $l$.

In addition to the CHPs, we also observed CPPs in the α-MoO$_3$/graphene heterostructure. Apart from the α-MoO$_3$ slabs, most of the heterostructure is thin-α-MoO$_3$-covered graphene (Figure 1b). Parallel polariton fringes were also observed close to the crystal edge, and they share similar features as CHPs on the α-MoO$_3$ slabs. We performed a similar analysis of these polaritons by extracting the line profiles (Figure 5a) and ω-$k$ dispersion (pink squares, Figure 5b) and comparing them with the simulation (false color, Figure 5b). A good fit between the experiment and calculation can be obtained by inputting $E_F$ = 0.6 eV for the graphene, consistent with our analysis for CHPs and the DFT results. Note that polaritons in thin-α-MoO$_3$-covered graphene mainly stem from the plasmonic responses of graphene. Thin α-MoO$_3$ facilitates the CT but merely affects the overall polaritonic responses (see Supplementary S2 for details). Therefore, polaritons outside the α-MoO$_3$ slabs are referred to as CPPs[13].

s-SNOM nano-imaging data augmented with electromagnetics and DFT calculations in Figures 1–5 demonstrate direct CT between polaritonic α-MoO$_3$ and graphene and CHPs in their heterostructures. The CT—not evident in mechanically assembled α-MoO$_3$/graphene heterostructures[13]—was uniquely facilitated by the relatively pristine and adsorbates-free vdW epitaxial heterointerfaces. The CHPs were observed at various hyperbolic dispersion branches and exhibited evident wavelength elongation, with $\Delta\lambda$ reaching 150%. Future works may be directed towards delicate growth or patterning[45] of the heterostructure to locally engineer CT polaritons for nano-optical circuits, computation, communication, light emission and beaming. It is also promising to explore altering the vacancy density by pressure, temperature, and intercalation treatment of the heterostructures to further control the CT and CT polaritons.

**Methods**
*Epitaxial growth of* **α-MoO$_3$/graphene** *heterostructures*

The α-MoO$_3$/graphene heterostructures were grown by adopting our previously reported methods[14]. As a source substrate, 50nm Mo film was deposited on the SiO$_2$ (285 nm)/Si substrate by DC magnetron sputter. The source substrate was then placed on a preheated heater at ~ 525 °C. The target substrate—exfoliated graphene on SiO$_2$/Si—was placed upside down with a distance of ~ 0.5 mm above the source substrate in ambient condition. After 10 minutes, the target substrate was removed from the source substrate and heater, and α-MoO$_3$ crystals of various thicknesses were epitaxially grown on the exfoliated graphene on SiO$_2$/Si.

*Raman spectroscopy*

The Raman spectra of α-MoO$_3$/graphene were acquired using a Raman spectroscopy (JASCO, NRS-4500) with a 532 nm laser. The polarized Raman spectra were measured with a parallel configuration of the incident and scattered light at a fixed angle. The α-MoO$_3$/graphene sample was rotated during the measurement.

*Infrared nano-imaging*

The CT hyperbolic polaritons in α-MoO$_3$/graphene heterostructures were characterized by infrared nano-imaging using the scattering-type scanning near-field optical microscope (s-SNOM). In this work, the s-SNOM is a commercial product from Attocube based on an atomic force microscope (AFM) operated in the tapping mode. The AFM tip, coated with PtIr and a radius of approximately 10 nm, was tapped at a frequency of ~ 280 kHz and an amplitude of ~ 70 nm. In the experiment, monochromatic mid-IR quantum cascade lasers (QCLs), spanning from 850 to 1750 cm$^{-1}$, were utilized to illuminate the AFM. In order to acquire the authentic near-field signal, the s-SNOM nano-images were recorded by the pseudo heterodyne interferometric method, where the scattered signal from the AFM tip was demodulated at the third harmonics of the tip tapping frequency.

*Density Function Theory (DFT) Calculations*

The DFT calculations were carried out in meta-generalized-gradient approximation (meta-GGA) with SCAN+rVV10 van der Waals density functional using VASP density functional code[46-48]. A plane-wave cutoff energy of 400 eV was used. Brillouin zone integrations used a 4×4×1 Monkhorst-Pack *k*-point sampling grid for geometry optimizations with 0.01 eV/Å forces convergence criteria. The tetrahedron method with Bloch corrections was used for DOS calculations to define Fermi level position properly with 30×22×1 mesh centered at the Gamma point. For charge transfer calculation 8×8×1 Monkhorst-Pack *k*-point sampling grid was used. α-MoO$_3$ surface was built from 6 layers of α-MoO$_3$ (48 α-MoO$_3$ atoms and 12 carbon atoms) and 15 Å vacuum gap. Bader charge analysis[49] was performed to calculate charge transfer from the graphene layer to the α-MoO$_3$ surface.

**Supplementary Material**

See the supplementary material for the Raman spectroscopy of α-MoO3/graphene heterostructures, modeling of energy-momentum (ω-*k*) dispersion of charge transfer polaritons in α-MoO$_3$/graphene heterostructures, and the emphasize of the effect of charge transfer (CT) on altering the ω-*k* dispersion of hyperbolic polaritons.

**Acknowledgments**

S.D. acknowledges the support from the National Science Foundation under Grant No. DMR-2238691, OIA-2033454, and ACS PRF fund 66229-DNI6. K.K. and V.K. acknowledge the Texas Advanced Computing Center (TACC) at The University of Texas at Austin for providing access to the Frontera computational cluster (allocation DMR22032) [https://doi.org/10.1145/3311790.3396656]. G.H.L acknowledges the support of the National Research Foundation (NRF) funded by the Korean Government (Nos. 2021R1A2C3014316 and 2017R1A5A1014862) (SRC program: vdWMRC center), the Creative-Pioneering Researchers Program, the Research Institute of Advanced Materials (RIAM), Institute of Engineering Research (IER), Institute of Applied Physics (IAP), and Inter-University Semiconductor Research Center (ISRC) at the Seoul National University. J.S. acknowledges financial support from the Alabama Graduate Research Scholars Program (GRSP) funded through the Alabama Commission for Higher Education and administered by the Alabama EPSCoR.

**Conflict of interest**
The authors have no conflicts to disclose.

**Data Availability**
The data that support the findings of this study are available from the corresponding authors upon reasonable request.

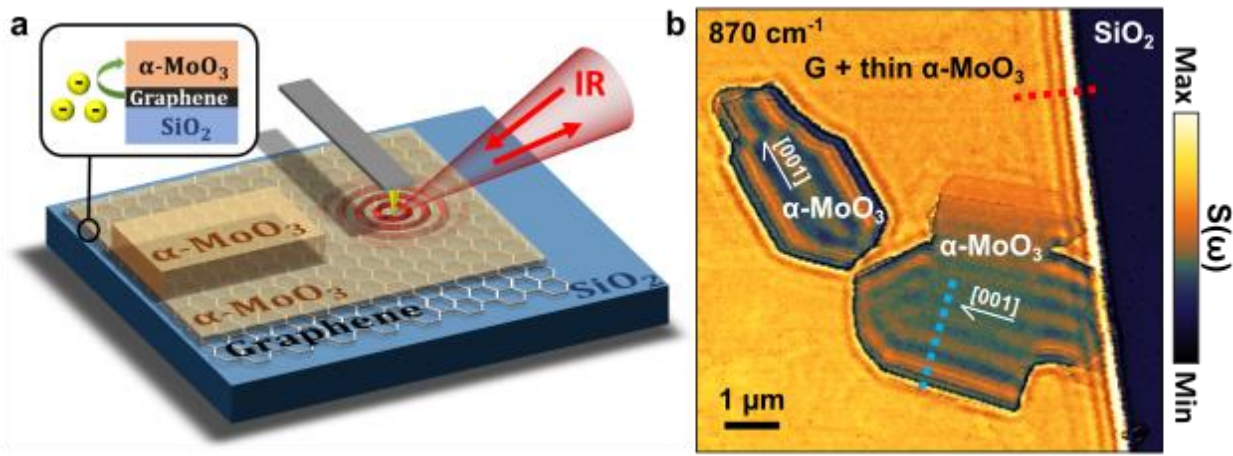

**Figure 1 | Scattering-type scanning near-field optical microscopy (s-SNOM) nano-imaging of charge transfer hyperbolic polaritons (CHPs) in the α-MoO₃/graphene heterostructure.** (a) Schematic of the s-SNOM investigation of CHPs in the α-MoO₃/graphene heterostructure. In the experiment, the infrared light (red arrows) was focused on the s-SNOM tip and launched propagating polaritons (red circles). The inset shows CT between graphene and α-MoO₃, where the electrons were moved from graphene to the top α-MoO₃. (b) s-SNOM amplitude (false color) image at frequency ω = 870 cm$^{-1}$. s-SNOM line profiles in Figures 2a and 5a are obtained from the line cuts marked with cyan and red dashed lines. The thicknesses of the α-MoO₃ slabs ([001] directions are marked): 43 nm (top) and 40 nm (bottom). The thickness of the graphene + thin α-MoO₃ (polycrystalline): 3 nm. The scale bar is 1 μm.

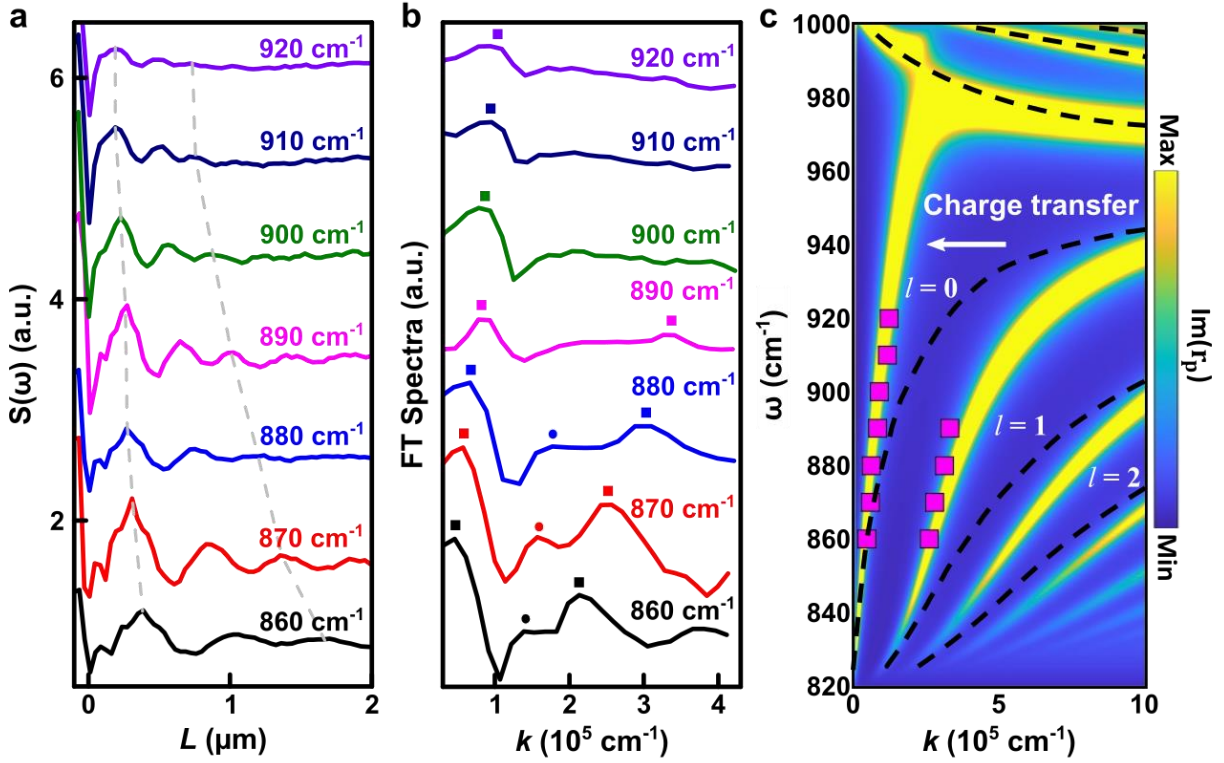

**Figure 2 | The scattering-type scanning near-field optical microscopy (s-SNOM) data of charge transfer hyperbolic polaritons (CHPs) and their energy-momentum (ω-k) dispersion.** (a) Line profiles along the cyan dotted line in Figure 1b at various frequencies ω. $L$ is the distance to the edge of the α-MoO$_3$ slab. The grey dashed curves mark the guide to the eye to reveal the dependence of CHPs on the frequency ω. (b) Fourier Transform (FT) Spectra of the s-SNOM line profiles in (a). The resonances of tip-launched CHPs are marked with squares, whereas edge-launched CHPs are marked with dots. (c) The ω-$k$ dispersion of CHPs. The theoretical results—Im($r_p$) of the α-MoO$_3$/graphene heterostructure—are plotted using the false color with the graphene $E_F$ = 0.6 eV. The dispersion branches of bare α-MoO$_3$ on SiO$_2$ are plotted with black dashed curves. The experimental data extracted from (b) are plotted with pink squares.

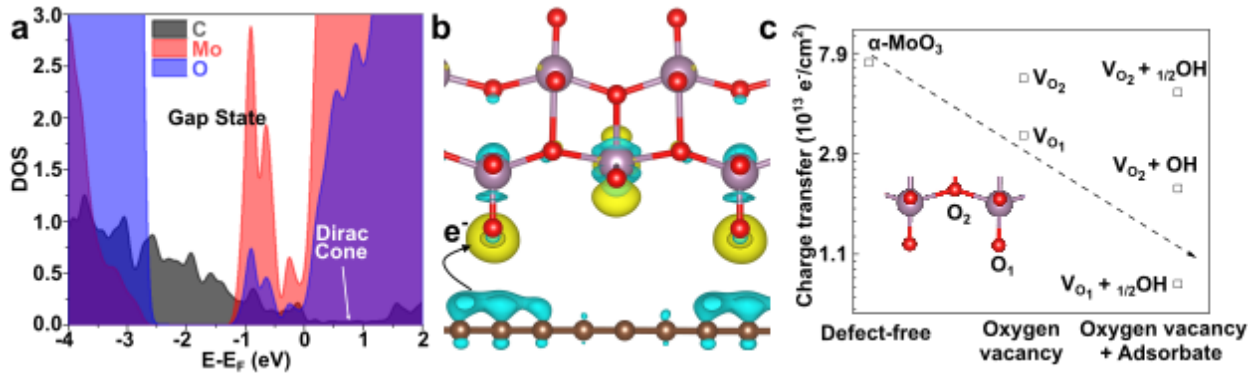

**Figure 3. The Density Functional Theory (DFT) calculation of charge transfer in the epitaxial α-MoO₃/graphene heterostructure.** (a) Projected density of states and (b) charge density difference for non-stichometry α-MoO$_3$/graphene heterostructure obtained using SCAN+rVV10 calculations. Yellow and cyan colors represent the charge accumulation and depletion regions, respectively. The iso-value is $9 \times 10^{-4}$. Fermi energy shift induced by charge transfer is 0.6 eV. (c) charge transfer for the α-MoO$_3$/graphene heterostructures as a function of surface oxygen vacancies or adsorbed moisture on the α-MoO$_3$ surface.

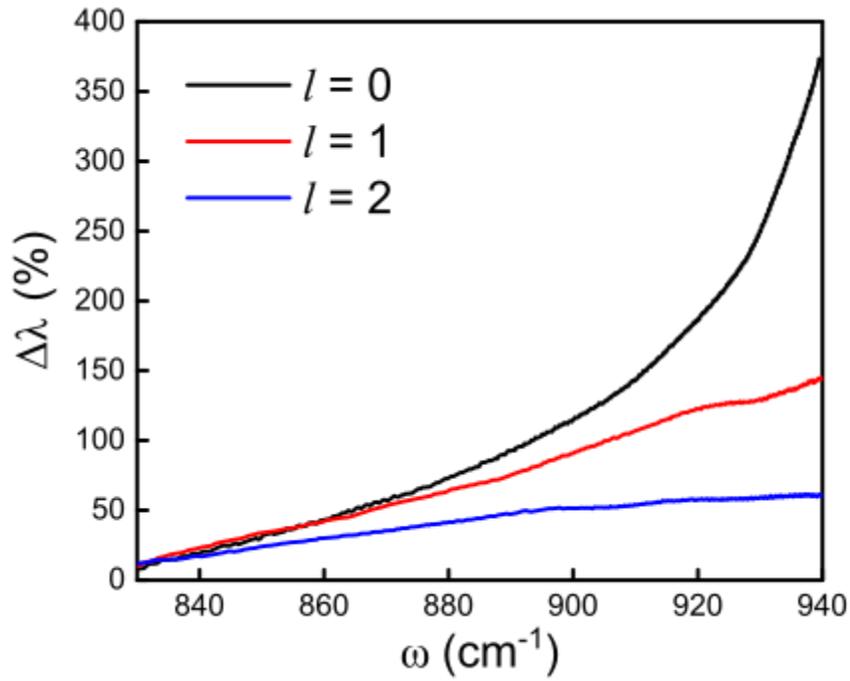

**Figure 4 | Charge-transfer hyperbolic polariton (CHP) wavelength elongation Δλ at various dispersion branches.** $l$ is the mode index defined in Figure 2. $\Delta\lambda = (\lambda_{CHP} - \lambda_{HP}) / \lambda_{HP}$, where $\lambda_{HP}$ is the wavelength of CHP in α-MoO$_3$-graphene heterostructure, and $\lambda_{HP}$ is the wavelength of hyperbolic polariton (HP) in bare α-MoO$_3$.

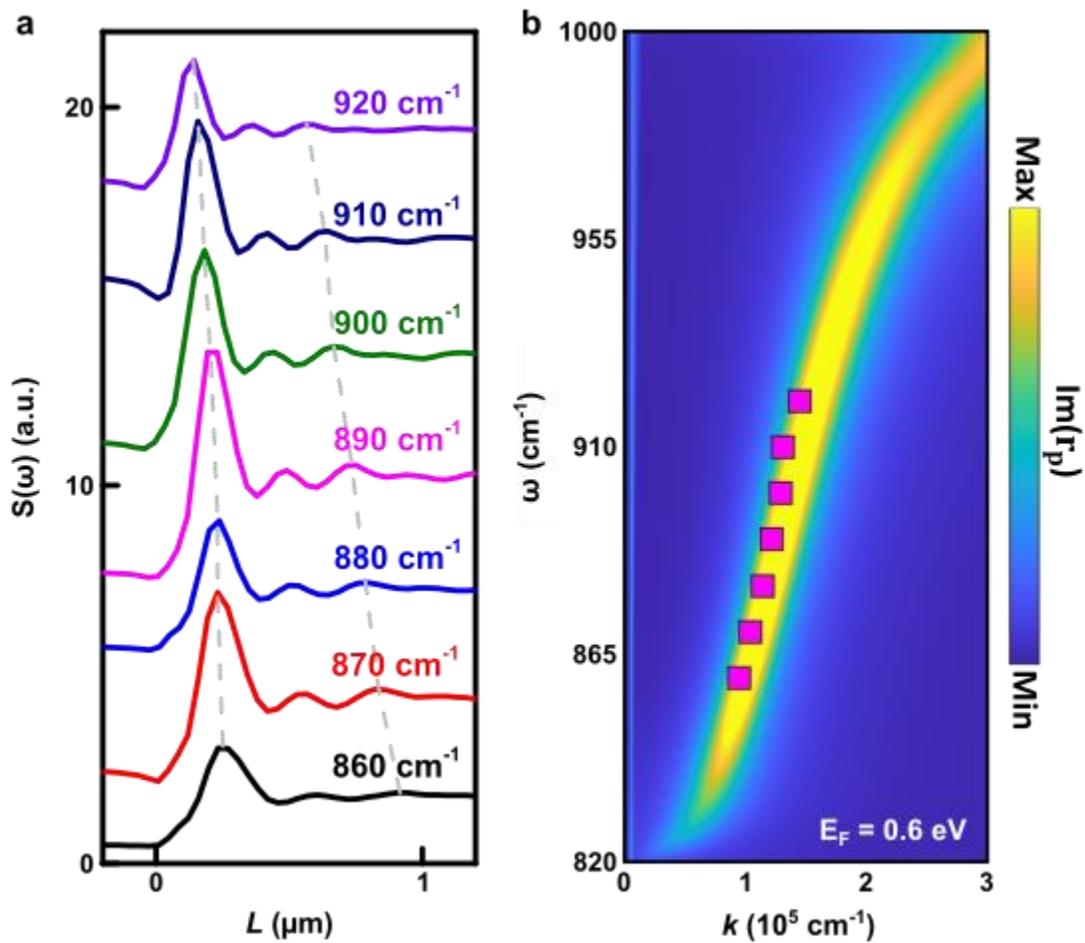

**Figure 5 | The scattering-type scanning near-field optical microscopy (s-SNOM) data of charge transfer plasmon polaritons (CPPs) and their energy-momentum (ω-k) dispersion.** (a) Line profiles along the red dotted line in Figure 1b at various frequencies. $L$ is the distance to the graphene edge. The grey dashed curves mark the guide to the eye to reveal the dependence of CPPs on the frequency ω. (b) The ω-$k$ dispersion of CPPs. The theoretical results—Im($r_p$) of the α-MoO$_3$/graphene heterostructure—are plotted using the false color with the graphene $E_f$ = 0.6 eV. The s-SNOM data are plotted with pink squares.